\documentclass[preprint,12pt]{elsarticle}

\usepackage{amssymb,eurosym}

\journal{IRBM}

\begin{document}

\begin{frontmatter}



\title{Ergonomic risk assessment of developing musculoskeletal disorders in workers with the Microsoft Kinect: TRACK TMS}


  \author{F. Buisseret\fnref{label1,label2}}
 \ead{buisseretf@helha.be}
  \author{F. Dierick\fnref{label2}}
 \ead{dierickf@helha.be}
 \author{O. Hamzaoui\fnref{label1}}
 \ead{oussama.hamzaoui@cerisic.be}
 \author{L. Jojczyk\fnref{label1}}
 \ead{jojczykl@helha.be}

\address[label1]{Centre d'Etudes et Recherche de la cat\'{e}gorie technique de la HELHa, Haute-Ecole Louvain en Hainaut (HELHa), Chauss\'{e}e de Binche 159, B-7000  Mons}
\address[label2]{Unit\'{e} de Recherche Forme et Fonctionnement Humain, Haute-Ecole Louvain en Hainaut (HELHa), Rue Trieu Kaisin 136, B-6061 Charleroi}



\begin{abstract}
Routine ergonomic assessment of postures and gestures in the workplace are mostly conducted by visual observations, either direct or based on video recordings. Nowadays, low-cost three-dimensional cameras like Microsoft Kinect open the possibility of recording the full kinematics of workers in a non-intrusive way, providing a more precise, and reliable assessment of their motor strategies. As an illustration, we focus on a peculiar kind of workers: professional musicians (violinists), whose playing is representative of a work situation involving repeated gestures and postures that can be described as non-ergonomic. We show that the Microsoft Kinect can be efficiently used to quantify the motion performed by these musicians. Moreover, we argue that low-cost three-dimensional cameras can be a useful aid in ergonomic risk assessment of developing musculoskeletal disorders and give the example of the repetition of movements and postural items included in the OCRA checklist, whose scoring can be facilitated by such a device, as addressed in our TRACK TMS research project. 
\end{abstract}

\begin{keyword}
3D kinematics; Kinect; Ergonomics; OCRA; Musicians


\end{keyword}

\end{frontmatter}


\section{Introduction}\label{sec1}

The huge expansion of video game industry has led to important technological advances in three-dimensional human motion tracking at an unbeatable price, one of the most popular example being the Microsoft Kinect v2 (MK). The existence of the Kinect SDK for Windows  allows the use of its kinematical data to other purposes than video games. In particular, several studies have shown the relevance of MK as an accurate and reliable tool for healthy and clinical assessment of gait and posture \cite{kinect1,kinect2,kinect3,kinect4,kinect5}. Because of its small size ($4.9$ cm $\times$ 6.6 cm $\times$ 6,7 cm), it is worth asking whether MK can also be an interesting tool for making direct observations in the workplace. In other words, may MK be a relevant device for ergonomic risk assessment of developing musculoskeletal disorders in workers \cite{tms}? The present proceeding aims at providing a positive answer to that question through some preliminary results of the TRACK TMS research project.

\section{Kinematics of professional violonists}\label{sec2}

We have developed a tracking application using the Kinect SDK for Windows in C$\sharp$, allowing the simultaneous recording of the three-dimensional coordinates of all the body points tracked by the MK. We reach a sampling frequency of 30 Hz. Simple tests performed in our laboratory (well-defined motions, like raising one's arm, drawing a square with one hand, etc.) allow to state that, with an optimal placement of the MK -- distance of 1 to 2 m in front of the subject -- , the accuracy of position tracking in a given workplace may be of order 1 cm. 

\begin{figure}
\includegraphics[scale=0.6]{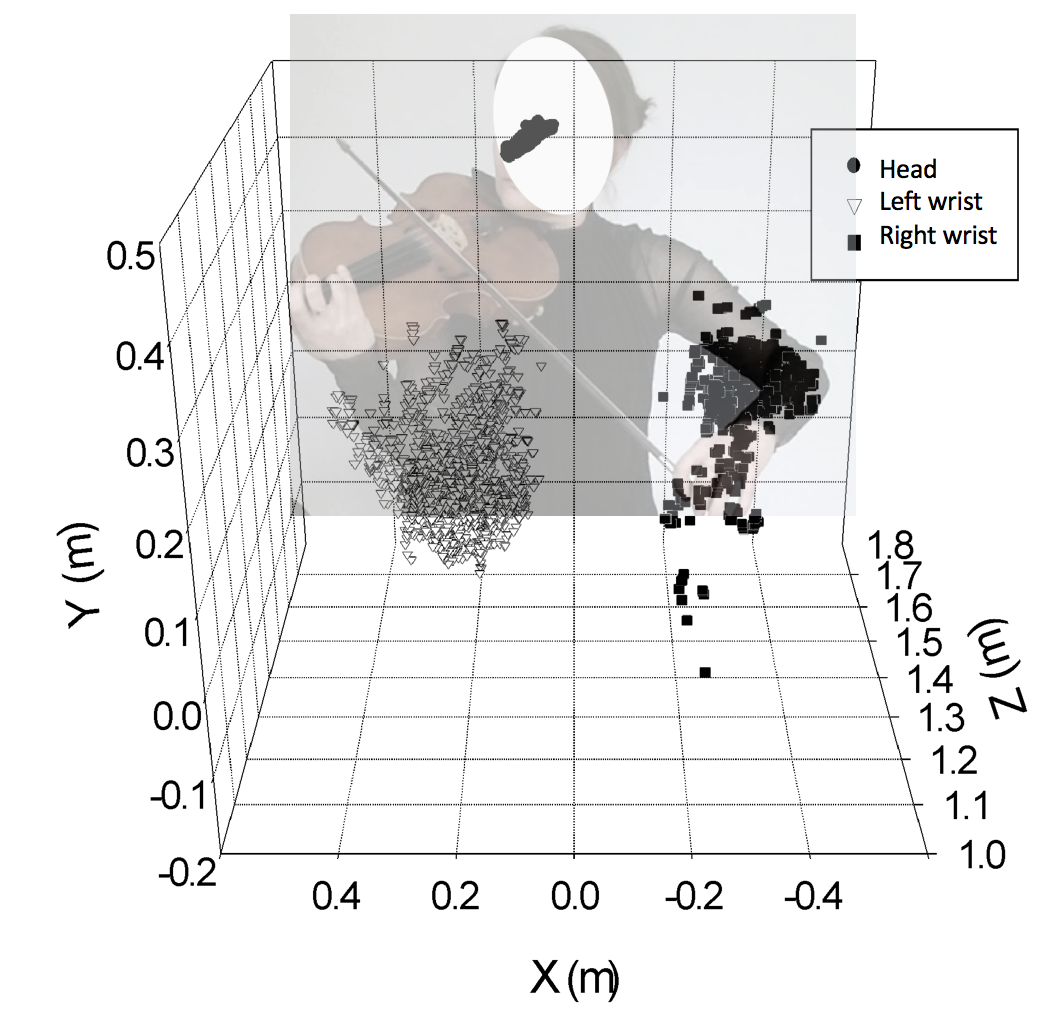}
\caption{Example of raw data acquisition by the Microsof Kinect: The positions of the head and the wrists of a violonist from the ORCW have been recorded during one minute of playing (points). A picture of the musician during the recording has been added for the sake of clarity. $X$ and $Y$ are the coordinates in the picture plane, and $Z$ is the depth, $Z=0$ being the position of the camera. The sampling frequency was 30 Hz. }
\label{Fig1}
\end{figure}

Among the great variety of workplaces worth to be studied, we have had the opportunity of collaborating with several musicians of the \textit{Orchestre Royal de Chambre de Wallonie} (ORCW, orcw.be). The development of musculoskeletal disorders among professional musicians are related to specific factors such as repeated gestures and non-ergonomic postures, both being often inherent to the practice of a musical instrument \cite{tm_mus2,tm_mus}. Note also that musicians are de facto handling an instrument or at least part of it (with the exception of pianists), which may be challenging for the MK since some body points may be from time to time hidden by the instrument itself. Checking the accuracy of motion tracking in the peculiar case of musicians is then somehow representative of other situations in which a worker is handling an object. A typical plot of the performed tracking is shown in Fig.~\ref{Fig1}, where we have plotted the violinist's head and wrists, giving a good representation of the whole motion. It appears that MK still correctly detects these points despite the presence of the bow and the handle.

\begin{figure}
\includegraphics[scale=0.6]{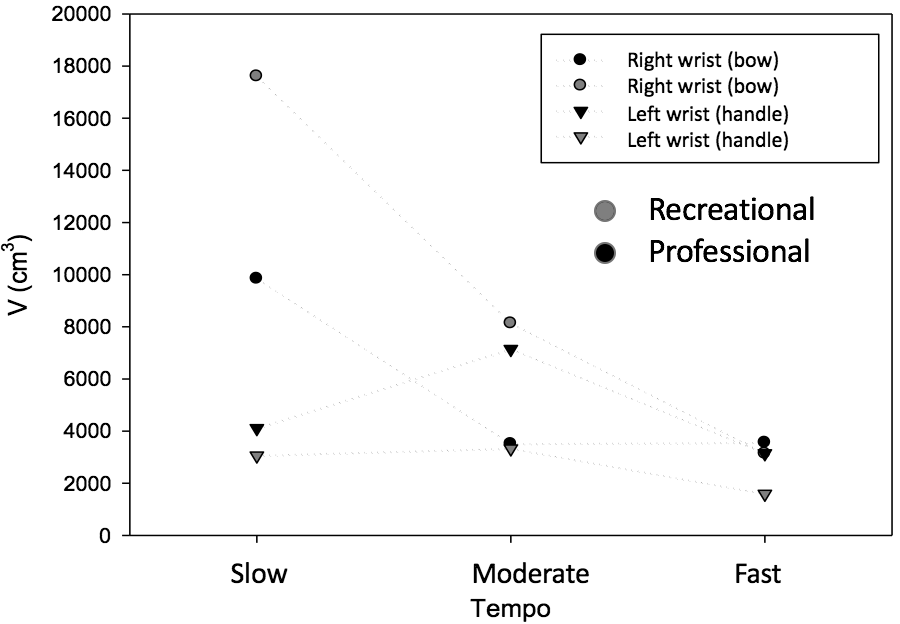}
\caption{Volume of the ellipsoid of confidence, computed from Eq. (\ref{volc}), for the left (triangles) and right (circles) wrists of a professional (black circles) and recreational (gray circles) violinists at three different tempi: Slow (around 60 BPM), Moderate (around 100 BPM) and Fast (above 140 BPM). Recording time was one minute for each tempo. }
\label{Fig2}
\end{figure}

A violinist playing his/her instrument is, from a kinematical point of view, fluctuating around a given, preferred position. For such a motion it can be assumed that the values of the time series $X^\alpha_i(t_n)$, giving the three coordinates ($i=1,2,3$) of the tracked point $\alpha$ at successive times $t_n$,  follow a normal distribution at sufficiently large $n$. A convenient way to quantify the motion is then to compute the volume of the ellipsoid of confidence $V^\alpha$, \textit{i.e.} the volume of the ellipsoid such that it contains 90$\%$ of the tracked coordinates $\{X^\alpha_1,X^\alpha_2,X^\alpha_3\}$ associated to the given body part $\alpha$. It is given by~\cite{regre} 
\begin{equation}\label{volc}
V^\alpha=\frac{4}{3}\pi \prod^3_{i=1}(1.954 \sigma^\alpha_i),
\end{equation}
$\sigma^\alpha_i$ being the standard deviation of the values $X^\alpha_i(t_n)$. As an illustration of the typical trends we observe, we have plotted in Fig.~\ref{Fig2} the volume of the ellipsoid of confidence, computed from Eq. (\ref{volc}), for the wrists of a professional violonist (from the ORCW) and a recreational one (less than 30 minutes of play per day)  at three different tempi. Subjects were actually asked to play one minute of a piece of their choice at the following tempi: lento, moderato, presto.  

The motor strategies observed are coherent with visual observation and can be commented as follows. For both musicians, playing at slow tempo is associated to large movements of the bow comparatively to the handle. On the contrary, playing at fast tempo consists in bow motions of smaller amplitude. At moderate tempo, the recreational musician adopts an intermediate strategy between the slow and fast tempi. The professional musician rather changes strategy and mostly moves the handle rather than the bow. This shows that three-dimensional kinematic coordinates of limited but nevertheless well chosen body points may give sufficient information about the strategies adopted by the musicians during their play. One musician's playing strategies are known to change with the level of experience and with the emotional state of the performer, as checked with a 8-camera system in \cite{music}. We see that, even with a single MK, such observations can in principle be made, although we focused on the tempo rather than on the emotional state. From an ergonomic perspective, such results, once confirmed by larger-scale studies could lead to practical suggestions such as varying the tempi during the rehearsals, for example. 

\section{Motion tracking and ergonomic assessment with OCRA}

Having in mind the above results suggesting that MK may be used as an efficient tool for motion tracking in musicians, it is worth wondering whether it could help to assess \textit{e.g.} the risk of developing musculoskeletal disorders associated to a peculiar workplace. We think the answer is also positive since assessment methods in ergonomics are often implicitly based on kinematics. Let us choose the example of the OCRA checklist \cite{ocra}. It mainly consists in scoring various items related to the actions performed by the worker: repetitiveness, muscular force, posture, and additional factors. We refer the interested reader to \cite{ocra} for a detailed description of these items. It is nevertheless worth giving three examples related to the repetitiveness and posture items: (1) How many actions per minutes are performed ? (2) What fraction of his/her time does the worker spends with the arm at the same height (or above) the shoulder ?
(3) What fraction of his/her time is the worker making large amplitude motions of the elbow/wrist/hand ?

The tracking of the body motion may answer unambiguously to these items in a more accurate way than with visual observation. Since such a tracking is available (see Fig. \ref{Fig1}) -- kinematics of the hand has to be excluded since MK is not designed to track fingers--, we therefore conclude that low-cost three-dimensional camera may become a reliable ``ergonomist assistant" in a near future \footnote{Reaching a sampling frequency of 30 Hz and a precision of order 1 cm on a tracked point is a priori accurate enough to score items in an ergonomical checklist such as the OCRA one.}. The implementation of a MK-based OCRA evaluation is currently in progress. As an example, let us show how any motion's repetitiveness can be evaluated from the recorded kinematics. Recent works have proposed powerful algorithms to identify the most relevant (carrying most of the information) degrees of freedom from arbitrary motions on the basis of MK kinematics \cite{motion1, motion2}. Once these degrees of freedom are identified, standard computations such as the power spectral density of the selected time series may straightforwardly give the motion's dominant frequency through the identification of the highest peak in the power spectral density. The number of actions per minute can then be estimated from the corresponding frequency, without a priori assumption on the periodic nature of the signal.

\begin{figure}
\includegraphics[scale=0.5]{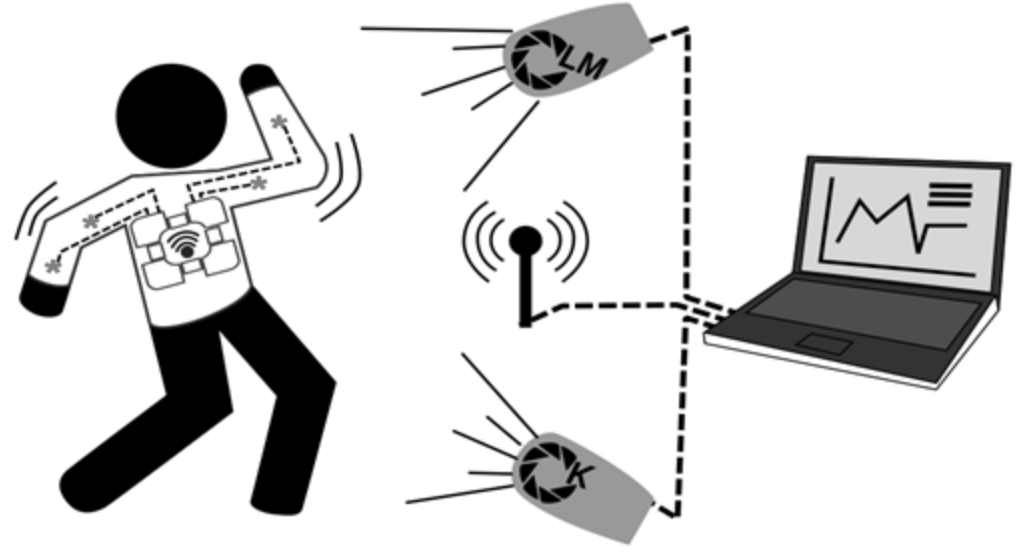}
\caption{Schematic representation of the TRACK TMS project. The device is made of three parts: a global three-dimensional camera (K for the Kinect), a local three-dimensional camera (LM stands for Leap Motion) and a wireless intelligent clothing made of electrodes measuring muscular activity. }
\label{Fig3}
\end{figure}

Finally, we mention that using MK as a tool for ergonomic assessment is one part of the so-called TRACK TMS research project illustrated in Fig. \ref{Fig3}, aiming at developing a non-intrusive, low-cost, measurement device in view of assessing a worker's risk of developing musculoskeletal disorders related to his/her professional activity. The device will be made of three parts: a global three-dimensional camera (MK), a local three-dimensional camera focused on hand kinematics (leapmotion.com), and home made electrodes measuring muscular activity. We let the presentation of the full project and a more complete study of its efficiency for future reporting.

\section*{Acknowledgements}
The authors thank L. Fack (ORCW's CEO) and the musicians of the ORCW, C. L\'{e}ger and B. Soyez (HELHa) , and Ph. Brux (MODYVA's CEO) for their spontaneous participation to this project. O. Hamzaoui acknowledges financial support from the First Haute-Ecole programme, project n$^\circ$ 1510470, TRACK TMS, in partnership with MODYVA (modyva.be).

\end{document}